\documentclass[12pt,twoside]{article}
\usepackage{fleqn,espcrc1}
\input epsf

\usepackage{graphicx}

\newcommand{\AmS}{{\protect\the\textfont2
  A\kern-.1667em\lower.5ex\hbox{M}\kern-.125emS}}

\title {Probing the Structure of Nucleons in the Resonance region with CLAS 
at Jefferson Lab}

\author{Volker D. Burkert\address{Thomas Jefferson National Accelerator 
Facility, \\ 
12000 Jefferson Avenue, 
 Newport News, Virginia, USA}}

\begin{document}

\maketitle

\vspace{-5.5cm}
\hspace{12cm}JLAB-PHY-00-10
\vspace{5.0cm}

\noindent
for the CLAS collaboration

\vspace{0.5cm}

\begin{abstract}
The physics of electromagnetic excitation of nucleon 
resonances, and their relevance in nucleon structure studies are 
discussed. Preliminary results from the CLAS detector at Jefferson Lab
are presented, and future prospects are discussed.
\end{abstract}

\section{INTRODUCTION}

Understanding the light-quark baryon spectrum and the electromagnetic 
transition amplitudes from the nucleon ground state to the excited states 
provides insight into the spatial structure and the 
spin structure of the nucleon. In this domain, constituent quarks and glue 
rather than elementary quarks and gluons appear to be the prevalent degrees 
of freedom. However, there is evidence that hadronic degrees-of-freedom are 
important as well, and there exists a well-known phenomenological connection 
to the valence quark regime in deep-inealstic scattering, known as 
Bloom-Gilman duality. In order to put the resonance region on the solid footing 
of QCD, the relative importance of these degrees 
of freedom as a function of the distance scales must be examined. 
Photo- and electroproduction of mesons from nucleons provide the most direct 
information about the spatial and spin structure of the excited states.       

The following are areas where the lack of high quality data is most 
noticeable, and where data from CLAS will contribute significantly.

\begin{itemize}

\item{To understand the internal nucleon structure, we need to study 
the full excitation 
spectrum as well as the continuum. While the continuum has been 
studied extensively, 
none of the resonance transitions have been studied well over a large enough 
distance scale.} 

\item{The known spectrum appears rather incomplete when compared with our
most accepted constituent quark models \cite{isgkar}. 
Many states are missing from the spectrum, and 
some masses of well known states are not well reproduced.}

\item {The role of the glue in the baryon excitation spectrum is completely unknown, 
although gluonic excitations of the nucleon are expected to be produced 
copiously \cite{isgur}, and predictions of hybrid baryon masses and quantum numbers 
are available from bag models \cite{gohaka}, QCD sum rules \cite{kissl}, 
and flux tube model \cite{page}  estimates.}  

\item{The nucleon spin structure has been explored for more than two decades 
at high energies. 
The nucleon resonance region which gives dominant contributions to the 
spin structure functions and sum rules at small $Q^2$ \cite{gdh}, and the 
transition to the deep inelastic regime have hardly been explored at all.}  
 
\item {The long-known connection between the deep inelastic regime and the 
resonance region (parton-hadron duality) \cite{blogil} 
remained virtually unexplored in its potential to obtain a better understanding of
the nucleon structure.}

\end{itemize}

\noindent All these topics are currently studied at JLAB, many employing 
the CLAS detector \cite{mecking}. 
I will focus exclusively on the resonance region and the first 
preliminary data from CLAS that begin to elucidate some of these aspects 
of nucleon structure.

\section{THE QUADRUPOLE TRANSITION TO THE $\Delta(1232)$}

The lowest excitation of the nucleon is the $\Delta(1232)$, the ground state of 
the isospin 3/2 spectrum. The 
electromagnetic excitation is dominanted by a quark spin flip corresponding
to a magnetic dipole transition $M_{1+}$. This contribution is well known 
up to fairly large $Q^2$.
The current interest is in probing the small electric ($E_{1+}$) and scalar quadrupole 
($S_{1+}$)
transitions. These are sensitive to possible deformations of 
the nucleon or the $\Delta(1232)$ from spherical symmetry.
Contributions at the few percent level to the ratios $R_{EM} = E_{1+}/M_{1+}$ 
and  $R_{SM} = S_{1+}/M_{1+}$  may result from interactions with 
the pion cloud at large and intermediate distances \cite{satlee,yang}. 
Quark models that include hyperfine interaction from one-gluon exchange 
predict small contributions as well \cite{iskako}.  
An intriguing prediction is that in the hard scattering limit the 
electric quadrupole contribution should be equal in strength to the 
magnetic dipole contribution \cite{carlson}. An analysis \cite{burelou} 
of earlier DESY data
found small nonzero values for $R_{EM}$ at $Q^2 = 3.2GeV^2$, 
showing that the hard scattering limit may be approached only at 
much larger values of $Q^2$ than currently accessible. 

A recent experiment at Jefferson Lab \cite{frolov} measured  
$p\pi^o$ production in the 
$\Delta(1232)$ 
region at high momentum transfer, and found values for
$E_{1+}/M_{1+} \approx -0.02$ at $Q^2 = 4~GeV^2$. The focus with CLAS is on the low to medium 
$Q^2$ regime, where data are sensitive to ingredients in nucleon structure models.
The $ep \rightarrow ep\pi^o$ is particularly suited to measure the $N\Delta$ transition multipoles. 
Complete distributions in the $p\pi^o$ azimuthal and cms polar angle have been measured over 
the hadronic mass range from threshold up to W = 1.6 GeV in a range in momentum transfer 
$Q^2 = 0.4~- 1.8$GeV$^2$.  
The data have been analysed using different approaches which give consistent 
results \cite{lcsmith,latifa}. Preliminary results for $R_{EM}$ and $R_{SM}$ are 
shown in Figure \ref{fig:e2m1} and \ref{fig:s2m1}, 
and compared with recent model calculations. $R_{EM}$ is 
small and negative, with a weak $Q^2$ dependence, while $R_{SM}$ exhibits a strong $Q^2$ 
dependence with a trend towards increasingly negative values. The trend of 
the data is 
qualitatively described by quark models that include pion degrees of freedom. 
These results are in contrast to previous data which show no clear $Q^2$ 
dependence of $R_{SM}$, and do give 
ambiguous results for the sign of $R_{EM}$.

\begin{figure}[tb]
\begin{minipage}{0.47\textwidth}
\epsfysize=7.9truecm
\epsfxsize=7.9truecm
\epsfbox{e2m1.epsi}
\caption{Preliminary CLAS results for $R_{EM}$ of the N$\Delta(1232)$ 
transition. The curves represent recent models within a constituent quark model
including mesons cloud effects \protect\cite{yang,satlee}, 
and a chiral quark soliton model 
\protect\cite{goeke}, respectively}
\label{fig:e2m1}
\end{minipage}
\begin{minipage}{0.47\textwidth}
\epsfysize=7.9truecm
\epsfxsize=7.9truecm
\epsfbox{s2m1.epsi}
\caption{Preliminary CLAS results for $R_{SM}$ of the N$\Delta(1232)$ 
transition. Same models as in Figure \ref{fig:e2m1}.
\vspace{1.3cm}}
\label{fig:s2m1}
\end{minipage}
\end{figure}

\section {\bf WHAT IS SO SPECIAL ABOUT THE ROPER RESONANCE?}

The  internal structure of the $N^*(1440)$ has been the subject of an 
intensive 
debate in 
recent years. It is clearly visible as a resonant state 
in $\pi N \rightarrow \pi N$ and 
$\gamma N \rightarrow \pi N$ 
scattering. However, its transition strength drops rapidly with $Q^2$ in 
electroproduction \cite{gerhardt}, and its longitudinal coupling
is weak. Neither of these properties is well described in non-relativistic
constituent quark models which assign the state to a radial 
excitation of the 
nucleon. Moreover, its mass is lower than most models would predict. 
Models trying to explain these features range from assigning a large gluonic 
component \cite{libuli}, using light-cone kinematics \cite{capkei}, 
including strong meson cloud effects \cite{cato,dong}, 
to describing it as a molecular-type bound system of a nucleon and a $\sigma$ 
pseudo-particle \cite{krewald}. A clear distinction between a gluonic model
for the Roper, and meson cloud models, as well as light-cone models, is that the
latter ones all predict a zero crossing of $A_{1/2}(Q^2)$ while the gluonic 
model does not. Moreover, the scalar amplitude $S_{1/2}$ should be 0 for a 
gluonic Roper, while it is large for other models.

\begin{figure}[tb]
\epsfysize=8.0cm
\epsfxsize=15.5cm
\epsfbox{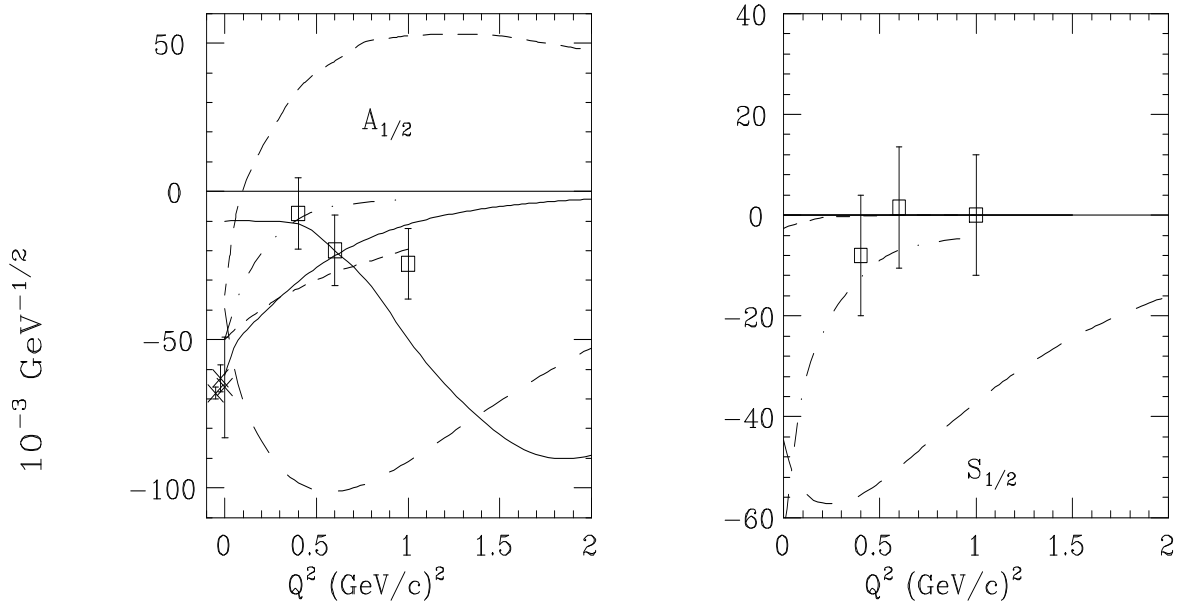}
\caption{Transition amplitudes $A_{1/2}$ and $S_{1/2}$ for the 
$\gamma_v pN(1440)$. The curves are from a quark model using light cone 
kinematics (dashed), a gluonic excitation model \protect\cite{libuli} (solid),
a non-relativistic dnamical quark model without (long dashes) and with 
(thin solid) relativistic corrections. The short dashed and dashed-dotted 
lines represent the boundaries of a fit to the data \cite{gerhardt}. }
\label{fig:roper}
\end{figure}

Although these models may qualitatively explain the fast drop of the 
transverse transition amplitude with $Q^2$, calculations exist only 
for the first two models. They make quite distinct predictions regarding
the $Q^2$ dependence as shown in Figure \ref{fig:roper}.
While recent flux tube model calculations \cite{page} 
give higher masses to gluonic (hybrid) baryons than previous estimates 
in the bag 
model \cite{gohaka} and QCD sum rules \cite{kissl}, the Roper could 
still have a substantial gluonic component due to mixing with higher 
mass gluonic states \cite{capstick}. 
Studying these transitions in electroexcitation allows us to probe 
the internal structure and reveal the true nature of this state.
Single $\pi^o$ and $\pi^+$ production as well as $N\pi\pi$ 
data from CLAS are currently being 
analyzed to determine the transition amplitudes in a large range of $Q^2$.

\begin{figure}[htbp]
\epsfysize=20.0truecm
\hoffset=3.5truecm
\epsfbox{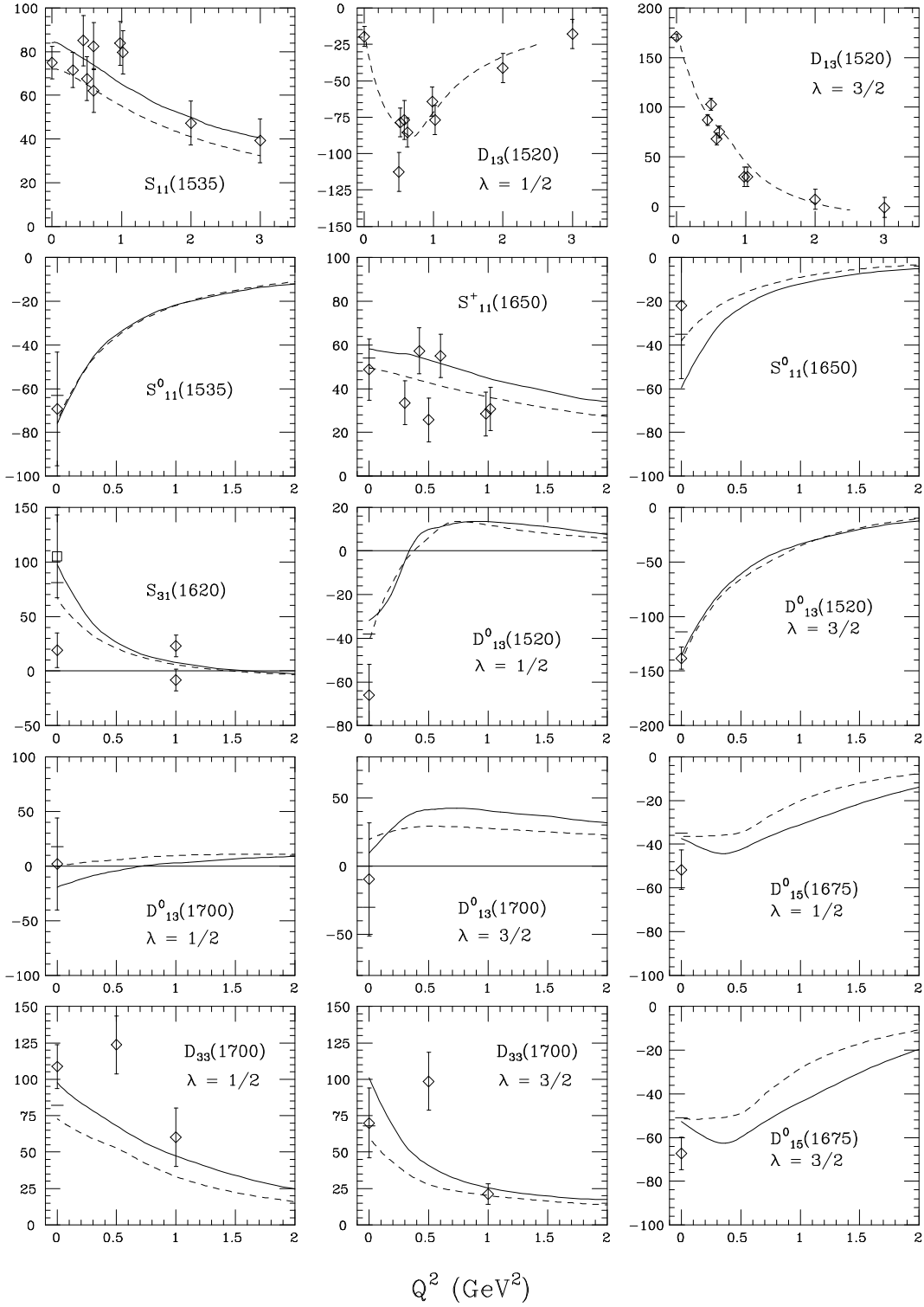}
\caption{Single Quark Transition Model predictions for states 
belonging to the $SU(6) \otimes O(3)$ multiplet, discussed in the text.}
\label{fig:sqtm}
\end{figure}

\section {\bf HIGHER MASS RESONANCES}

The total photoabsorption cross section shows only 3 or 4 enhancements; 
however, 
more than 20 states are known in the mass region up to 2 GeV. 
By measuring the electromagnetic transition of many of these 
states we obtain a  more complete picture of the 
nucleon structure, and provide the basis for testing symmetry 
properties of resonance transitions and the underlying 
3-quark symmetry group structure. For example, the approximate SU(6) 
symmetry of the non-relativistic symmetric quark model predicts 
relationships between transition amplitudes of  
states belonging to the same $SU(6)\otimes O(3)$ supermultiplets. 
For example, in the single-quark-transition model (SQTM), 
only one quark participates in the interaction. The model predicts 
transition amplitudes 
for a large number of states based on a few measured amplitudes 
\cite{hey}. At the photon point the symmetry relations are in good 
agreement with the data, showing that symmetry properties 
dominate over differences in the internal 
quark-gluon structure of different states. If the nucleon is probed 
at smaller distances, we expect 
symmetry properties to become less important.

The current situation is shown in Figure \ref{fig:sqtm}, 
where the SQTM amplitudes for 
the transition to the $[70,1^-]$ supermultiplet have been 
extracted from
the measured amplitudes for $S_{11}(1535)$ and $D_{13}(1520)$. Predictions 
for other
states belonging to the same multiplet are shown in the other panels 
\cite{prag}. 
The lack of accurate data for most resonances prevents a sensitive
test of the SQTM for space-like photons.

\begin{figure}[t]
\begin{minipage}{0.48\textwidth}
\epsfysize=10.5truecm
\epsfxsize=7.9truecm
\epsfbox{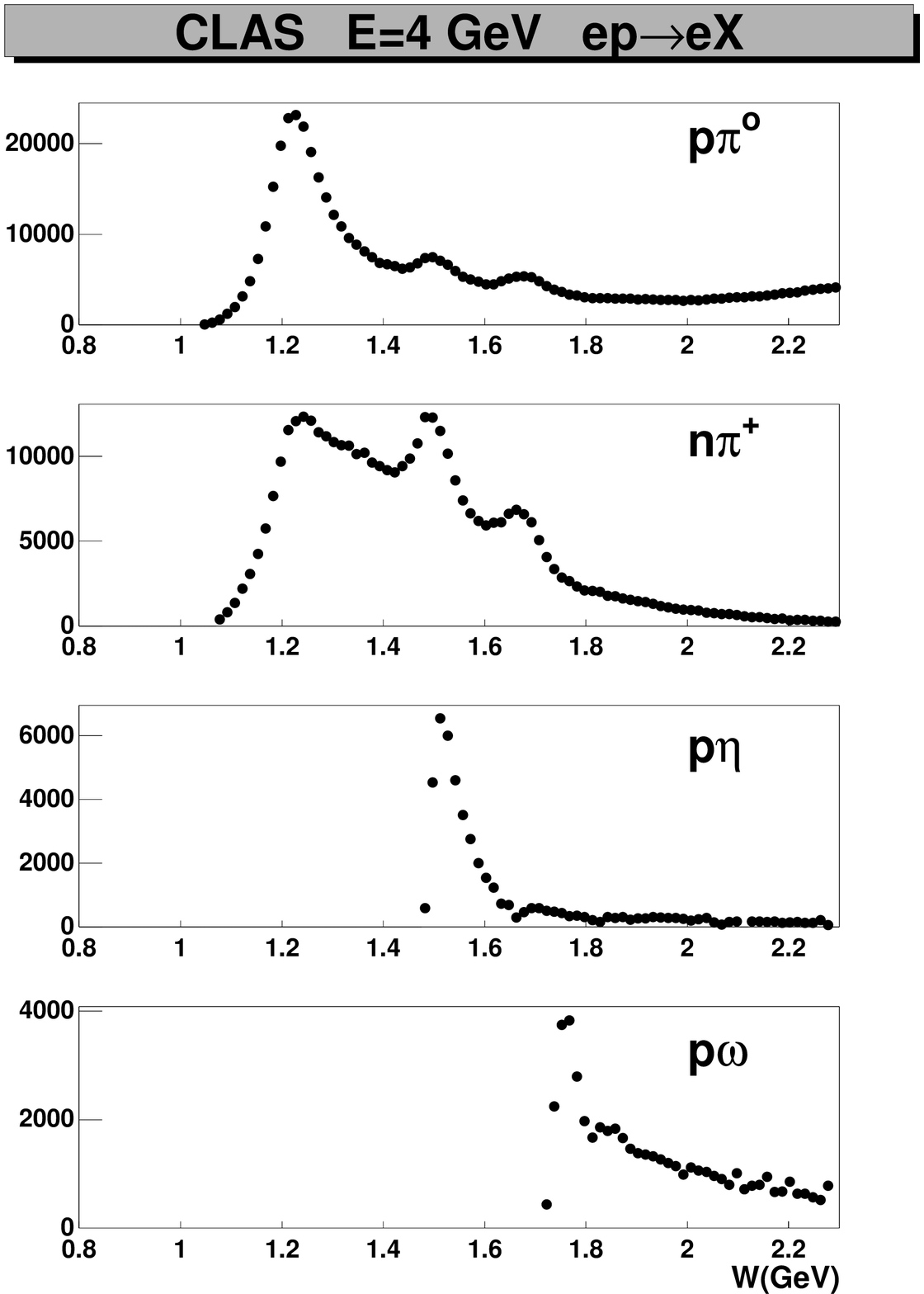}
\caption{\small Yields for various channels measured with CLAS at JLAB. 
The statistical error bars are smaller than the data points.}
\label{fig:yields}
\end{minipage}
\begin{minipage}{0.48\textwidth}
\epsfysize=10.5truecm
\epsfxsize=7.9truecm
\epsfbox{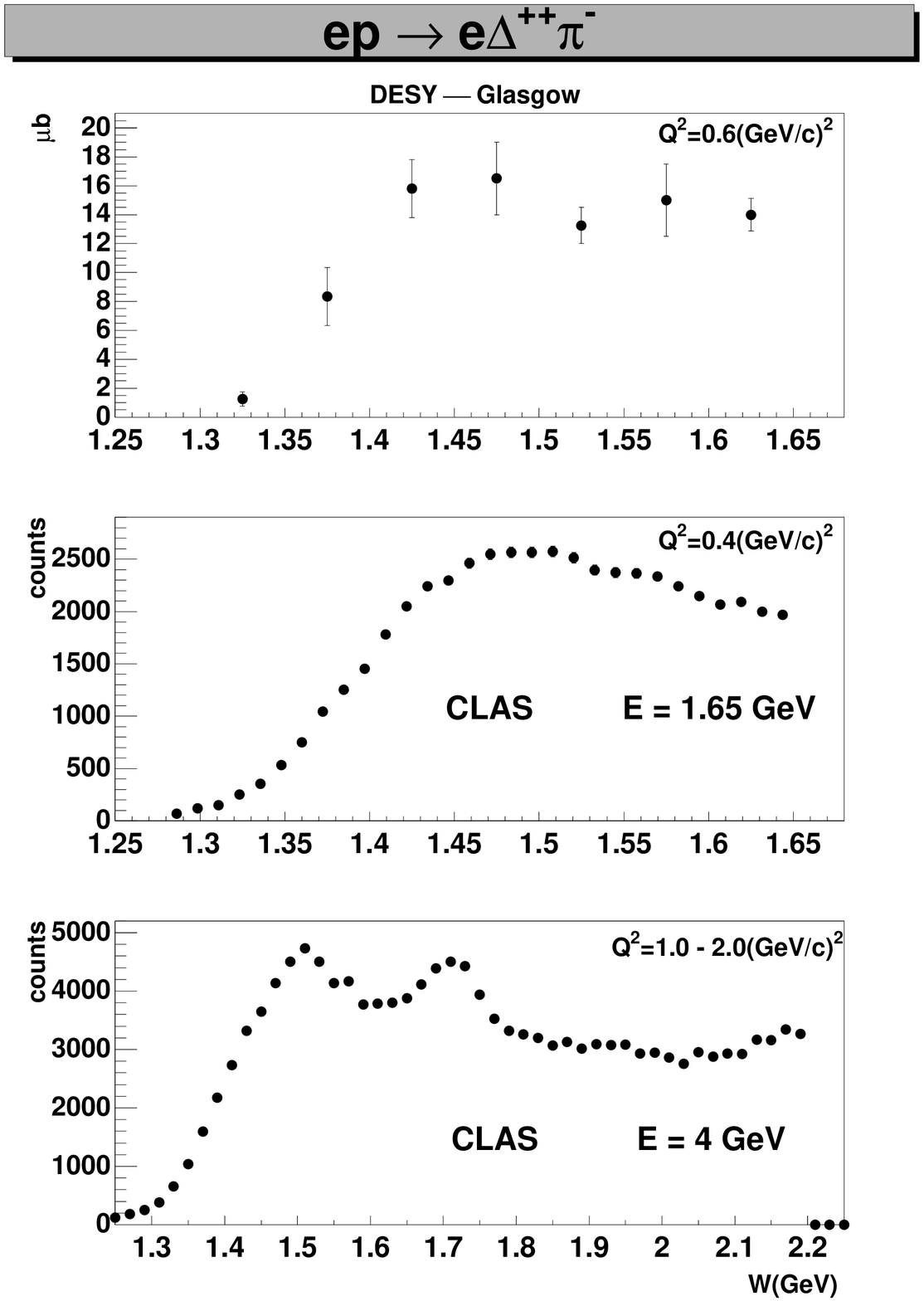}
\caption{\small Yields for the channel $\Delta^{++}\pi^-$ 
measured with CLAS at 
different $Q^2$ compared to previous data from DESY.}
\label{fig:deltapi}
\end{minipage}
\end{figure}

The goal of the experimental N* program at JLAB with the CLAS detector is to 
provide data in the entire resonance region, by measuring many
channels in a large kinematic range, including various polarization observables. 
The yields of several channels, recorded simultaneously in CLAS, are shown in 
Figure \ref{fig:yields}
and Figure \ref{fig:deltapi} . Resonance excitations seem to be present in all channels.
The graphs also illustrate how the various channels have sensitivity 
to different resonance excitations. For example, the $\Delta^{++}\pi^-$ channel 
clearly shows resonance excitation near 1720 MeV while single pion 
production is more sensitive to a resonance near 1680 MeV \cite{ripani}. 
The $p\omega$ channel seems to show resonance excitation near threshold, 
similar to the $p\eta$ channel. No resonance has been observed in this channel 
so far. 
The single pion channels are essential for an accurate determination of 
many transition amplitudes. For the first time,  $n\pi^+$ 
electroproduction has been measured throughout the resonance region, 
and in a nearly complete angle range.

\begin{figure}[tb]
\epsfysize=11.5truecm
\epsfbox{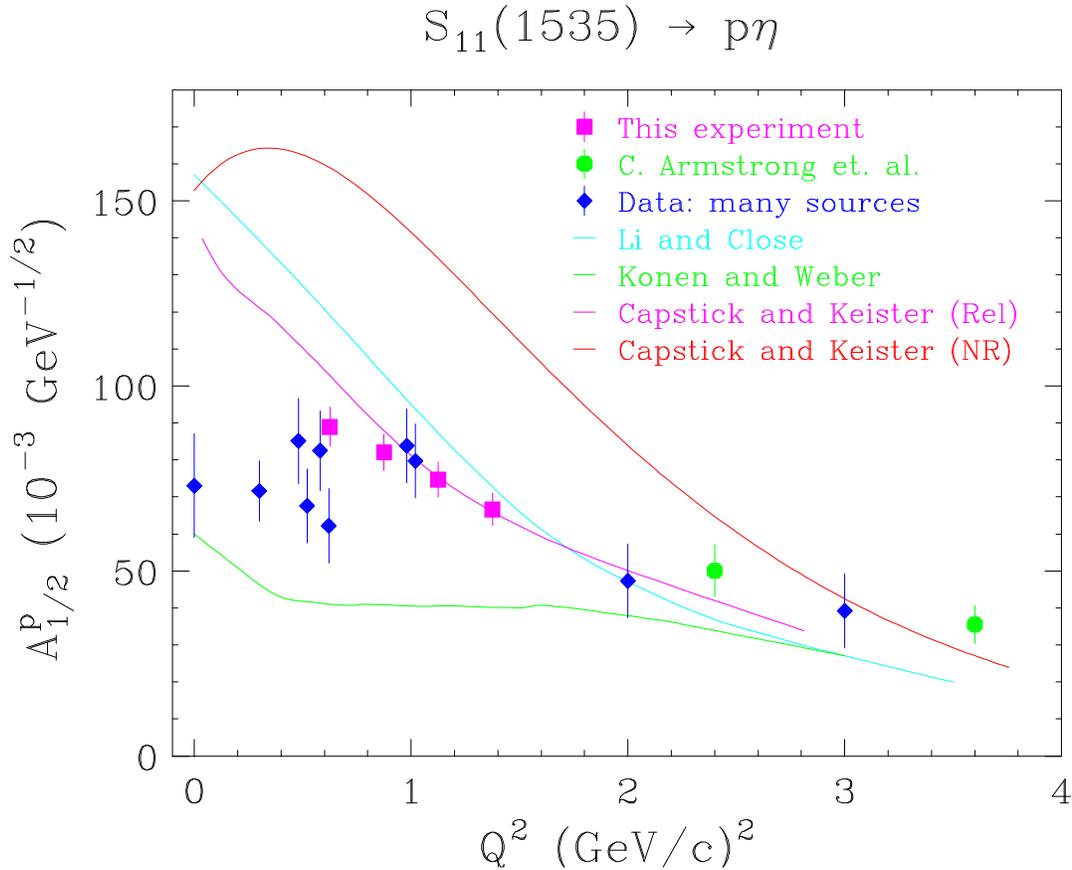}
\caption{\small Transverse photocoupling amplitude for the 
$\gamma_v p N^*(1535)$ transition. The full squares at lower $Q^2$ are 
preliminary CLAS data \protect\cite{dytman}. The full circles 
at large $Q^2$ are data from a previous JLAB experiment \protect\cite{armstrong}}
\label{fig:s11}
\end{figure}

New data have been obtained in the channel $ep \rightarrow ep\eta$. 
The $p\eta$ channel selects isopin 1/2 resonances, and is particularly 
sensitive to the $N^*(1535)$ state. Preliminary data from CLAS 
are shown in Figure \ref{fig:s11}. The data confirm the previously observed, 
and not fully 
understood, slow fall-off of the resonance transition form factor 
with $Q^2$. However, at low $Q^2$, the trend of the data favors 
a larger 
photocoupling amplitude for that state than the $Q^2 = 0$ data
point indicates. One should note 
that meson cloud effects 
can give different results for $N\pi$ or $N\eta$ channels unless 
the analysis properly accounts for rescattering and coupled 
channel effects.

For the first time the $N\pi\pi$ channel has  been measured with 
high statistics in photo-and electroproduction. Using an isobar model 
approach one can study contributions of $N\rho$ and $\Delta\pi$ decay 
channels \cite{ripani} which are especially sensitive to the so-called 
missing resonances. Figure \ref{fig:deltapi} illustrates  the vast improvement in data volume for the $\Delta^{++}\pi^-$ channel. 
The top panel shows DESY data taken more than 20 years ago.  The other 
two panels show samples of the data taken so far with CLAS. At higher $Q^2$, 
resonance structures, not seen before in this channel, are revealed.

\begin{figure}[tb]
\epsfysize=13.0cm
\epsfxsize=15.5cm
\includegraphics[angle=-90,totalheight=12.cm]{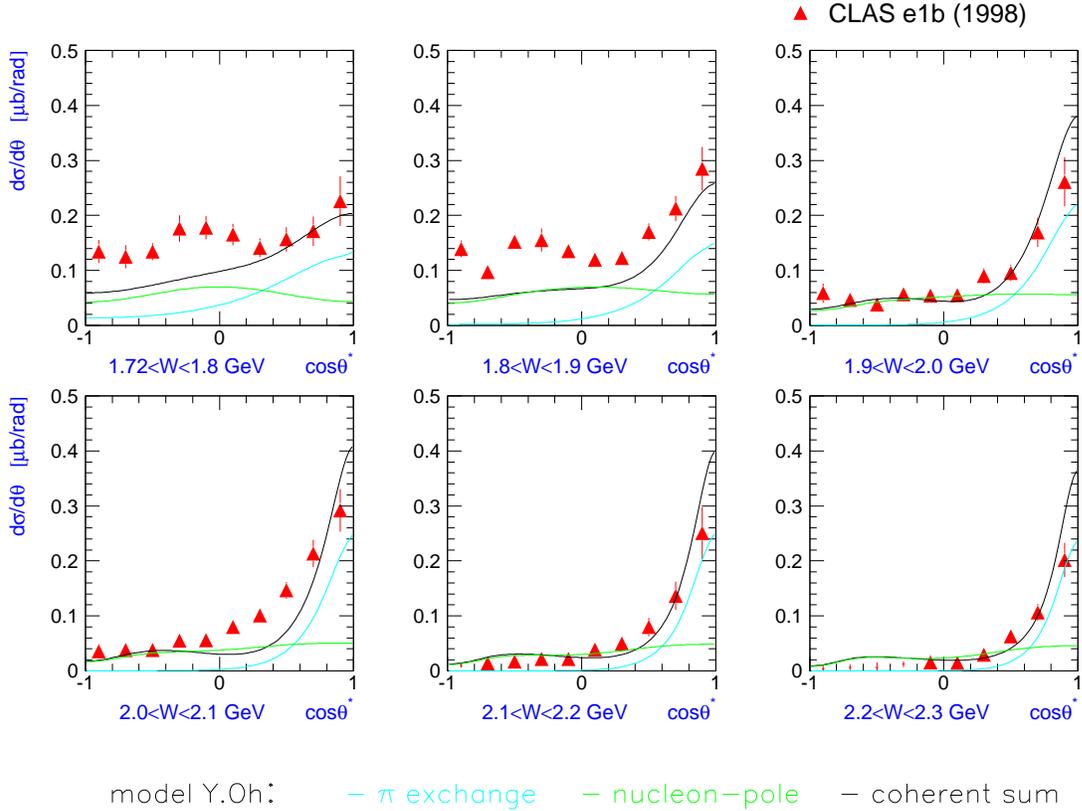}
\caption{Electroproduction of $\omega$ mesons for different W bins, 
and in a $Q^2$ range $\approx$ 1.0 - 2.0 GeV$^2$ 
\cite{fklein}.
The deviation 
of the $\cos\theta$ -distribution from a smooth fall-off for the 
low W bin suggests significant 
s-channel resonance production. Model calculation without resonance 
contributions \protect\cite{oh1}.}
\label{fig:omega}
\end{figure}

\section {\bf MISSING QUARK RESONANCES} 

With large acceptance detectors such as CLAS more complex final states can be studied 
than has been possible in the past. This will allow us to systematically tackle a 
longstanding problem, the so-called ``missing resonances''. 
These are states predicted in the symmetric $|Q^3>$ model to populate the 
mass region around 2 GeV but have
not been seen in $\pi N$ elastic scattering, which is 
our main source of information 
on the nucleon excitation spectrum. These states may thus be either absent 
from the spectrum, or they may not couple to the $N\pi$ channel.
It is important to search for at least some of these states since
their absence from the spectrum would be evidence that SU(6) symmetry 
is strongly violated in light-quark baryon spectroscopy.
Other symmetry scheme have been considered. For example,
the observed clustering of baryon states may reflect an underlying Lorentz-isospin
group symmetry $O(1,3)\otimes SU(2)_I$, as discussed recently in
ref. \cite{kirchbach}. The number of states is significantly reduced in this
latter approach compared to the symmetric model, more in accordance with
the observed spectrum. As these possibilities reflect very different
underlying baryon structure models, it is extremely important to search for 
states that are predicted in one scheme but not in the other. 

How do we search for states that do not couple to the $\pi N$ channel?
Channels which are predicted to couple strongly to ``missing'' states are
$N(\rho, \omega)$ or $\Delta\pi$. Some may also couple to $KY$ or 
$p\eta^{\prime}$ \cite{caprob}. Some of the ``missing'' states may also  
couple to $N\gamma$, and should therefore be excited in photo- or  
electroproduction experiments.

Figure \ref{fig:omega} shows very preliminary data from CLAS in the $p\omega$ 
channel. The 
process is expected to be dominated by t-channel $\pi^o$ 
exchange with strong peaking at forward $\omega$ angles, or low t,  
and a monotonic fall-off at large t. 
The data show clear deviations 
from the smooth fall-off for the W range near 1.9 GeV, where some of the 
``missing'' resonances are predicted, in comparison with the high W region. 
Although indications for resonance production are strong,
analysis of more data and a partial wave study are needed before 
definite conclusions can be drawn. 

CLAS has collected ~$3\cdot 10^5$ $p\eta^{\prime}$ events
in photoproduction. Production of $\eta^{\prime}$ has 
also been observed in electron scattering for the first time with CLAS. 
This channel may provide a new tool in the search for missing states
as well \cite{ritchie}.
The quark model predicts two resonances in this mass range 
with significant coupling to the $N\eta^{\prime}$ channel \cite{caprob}. 

$K\Lambda$ or $K\Sigma$ production may yet be another source of information on resonant
 states. Previous data show some evidence for resonance production in these 
channels \cite{bennhold}.
New data with much higher statistics are being accumulated with the CLAS
detector, both in photo- and electroproduction. Analysis
of the $\Lambda$ polarization provides additional information
sensitive to resonance excitations.

\section{OUTLOOK}

The experimental effort with CLAS at Jefferson Lab will provide the 
community with a wealth of data in the first decade of this new 
millennium to address many open problems in hadronic structure
in the domain of nucleon resonances and at intermediate distances. 
{\it The experimental effort must be accompanied by a significant theoretical 
effort to translate this into real progress in our understanding of the 
complex regime of strong interaction physics.} 
The region of nucleon resonances is of special interest as it represents a 
domain where different degrees of freedom, from hadronic, to 
constituent quarks, to valence quarks, overlap. On the one hand, this 
provides a challenge to theory, on the other hand an opportunity, as only 
under such circumstances can there be a realistic possibility for a unified 
description of hadron structure from small to large distances.

\end{document}